# Visualizing the Nonlinear Coupling between Strain and Electronic Nematicity in the Iron Pnictides by Elasto-Scanning Tunneling Spectroscopy

Authors: Erick F. Andrade[1*], Ayelet Notis Berger[1*], Ethan P. Rosenthal[1], Xiaoyu Wang[2], Lingyi Xing[3], Xiancheng Wang[3], Changqing Jin[3], Rafael M. Fernandes[2], Andrew J. Millis[1], Abhay N. Pasupathy[1]

Affiliations: [1]Department of Physics, Columbia University, New York, NY 10027, USA. [2]School of Physics and Astronomy, University of Minnesota, Minneapolis, MN 55455, USA [3] Institute of Physics, Chinese Academy of Sciences; School of Physical Sciences, University of Chinese Academy of Sciences; Collaborative Innovation Center of Quantum Matter, Beijing 100190, China.
* These authors contributed equally to this work

Abstract:
*Mechanical strain is a powerful technique for tuning electronic structure and interactions in quantum materials. In a system with tetragonal symmetry, a tunable uniaxial in-plane strain can be used to probe nematic correlations in the same way that a tunable magnetic field is used to probe magnetic correlations. Here, we present a new spectroscopic scanned probe technique that provides atomic-resolution insight into the effect of anisotropic strain on the electronic structure. We use this technique to study nematic fluctuations and nematic order across the phase diagram of a prototypical iron-based superconductor. By extracting quantitatively the electronic anisotropy as function of applied strain, we show that while true long range nematic order is established at the tetragonal to orthorhombic structural transition temperature, sizable nematic fluctuations persist to high temperatures and also to the overdoped end of the superconducting dome. Remarkably, we find that uniaxial strain in the pnictides significantly enhances the amplitude of the nematic fluctuations, indicating a strong nonlinear coupling between structure and electronic nematicity.*

Due to its strong effect on energy gaps, effective masses and scattering rates, strain has been used extensively to tune the electronic properties of semiconductors. Less explored but equally interesting is the effect of strain on metallic or semimetallic materials with strong electronic interactions. Such materials often exhibit electronic nematic phases that spontaneously break a discrete rotational symmetry of the lattice[1] and as a consequence couple linearly to uniaxial strain. Strain thus offers a unique tuning knob to study nematic correlations in materials, analogous to the effect of magnetic fields on a ferromagnet. Recently, elegant measurements of the strain dependence of dc transport [2-4] and optical[5, 6] conductivity have provided new fundamental insights into electronic nematicity. However, it is desirable to move beyond the spatially averaged information provided by dc transport and optical spectroscopy and obtain local, spectroscopic insight into the effects of strain on quantum materials. In this work we present a new technique in which atomic resolution scanning tunneling microscopy and spectroscopy measurements are performed on samples subject to in-situ and continuously variable mechanical uniaxial strain. This opens a new experimental window for understanding and manipulating electronic nematic phenomena in quantum materials at the microscopic level.

Electronic nematicity is the spontaneous breaking of a rotational symmetry of an electron gas. Although not necessarily driven by the lattice, electronic nematic order couples linearly to certain elastic modes and, consequently, triggers a simultaneous structural transition. Nematic effects are observed in many quantum materials including 2D quantum hall states[7, 8], ruthenates[9], cuprates[10-15] and the iron-based superconductors[2, 3, 16-23], which we focus on here. Strong electronic nematicity is seen

across the 122[2, 3, 5, 16, 18, 21, 24-27], 111[19, 22, 28], 1111[26] and 11[29-31] families of iron-pnictide superconductors. The nematic electronic order is accompanied by a lowering of lattice symmetry from tetragonal to orthorhombic where the lattice elongates along one nearest-neighbor Fe-Fe bond (*a* axis) and compresses along the orthogonal direction (*b* axis)[32]. This transition occurs at a temperature $T_S$ and is followed by a transition to "stripe" uniaxial antiferromagnetic order at a lower temperature $T_{SDW}$. In transport, nematicity is manifested as a temperature and strain dependent resistivity anisotropy[2, 3]. In scanning tunneling microscopy (STM) measurements, electronic nematicity is visualized as two-fold ($C_2$) symmetric patterns in the local electronic density of states (LDOS) [16, 19, 20, 25, 31, 33, 34]. These patterns in the LDOS have length scales of several lattice constants, and are observed for energies within ~100meV of the Fermi level in several different families of iron-based superconductors.

While the existence of nematic order in the iron-arsenide materials has been convincingly demonstrated, key physics issues remain widely debated. A central question is about the origin of the nematicity, and its relevance to superconductivity and other phenomena.[35]. The absence of magnetic ordering in the nematic compound FeSe has been invoked as an argument against magnetically-driven nematic order[30, 36] and in favor of orbital-driven nematicity[37-39]. On the other hand, the disappearance of nematic order in several hole-doped $BaFe_2As_2$ compounds as the magnetic ground state changes from stripe to double-**Q**[40-42] has been interpreted as evidence that magnetism is essential to promote nematic order[35, 43-47] . However, because the vast majority of pnictide superconductor families display both an orthorhombic phase and stripe magnetism, direct experimental evidence as to whether orbital or spin fluctuations drive the transition remains scarce. Given that the energy scales associated with nematicity are close to, but several times larger than the superconducting gap[5, 18, 19], settling this question is of great importance to the physics of the pnictides.

The second major issue is on the role of the lattice to nematicity. While symmetry arguments prescribe the form of the coupling of nematic order and fluctuations to applied strain, the magnitude and impact of the nematic-strain coupling remains unknown. This coupling is particularly important to elucidate unusual effects observed in the "para-nematic" state existing above the nematic transition temperature $T_S$, which vanishes near optimal doping. Transport measurements[2, 3, 27, 48], nuclear magnetic resonance [49], optical data [50], and our previous STM measurements[19] reveal evidence of electronic nematicity at temperatures well above $T_S$. Two compelling scenarios, with very different implications for the nature of the normal state from which superconductivity arises, have been advanced: in the first, true electronic nematic order is established at a temperature $T^*>T_S$[48]. In this scenario, the transition at $T_S$ is not a true nematic phase transition, but rather, a meta-nematic transition at which the nematic order parameter increases from a small to a larger value and the associated lattice distortion becomes observable. An alternative explanation supported by elasto-resistance measurements[2, 3] is that true long-range nematic order is only established at $T_S$, but that strong nematic fluctuations persist up to much higher temperatures. In this scenario, the nematicity seen in STM experiments at high temperature arises from nematic fluctuations coupled to a symmetry-breaking field. The symmetry breaking would most likely be due to anisotropic strain arising from defects in the crystal structure and/or anisotropic differential thermal expansion with respect to the substrate to which the crystal is glued during measurement [51]. Systematic studies of local spectroscopic properties as a function of applied strain can help us answer these questions and gain invaluable information about the nature and impact of the nematic degrees of freedom and their coupling to strain.

In order to probe the relationship between mechanical strain and microscopic electronic nematicity, we designed an apparatus by which anisotropic mechanical strain can be continuously applied to a sample while atomically-resolved STM and STS (scanning tunneling spectroscopy) measurements are performed on the same area. We term the new technique Elasto-Scanning Tunneling Spectroscopy (or E-STS). The chief technical problems to overcome are the incorporation of the strain-producing apparatus within the available sample space of a few millimeters typical for high-resolution cryogenic STMs; the design of the experiment to allow the same nanoscopic area of the sample to be traced while varying the strain; and the typical issues associated with multilayer piezoelectrics including drift, creep and noise, especially at high temperature. Our design is shown in figure 1f and consists of a multilayer piezo actuator that expands/contracts along one axis by up to ±0.1% (which is the typical orthorhombic distortion of parent pnictide compounds) upon application of voltages ($V_{strain}$) of ±250 V. The single crystal sample is glued to the top face of the piezo actuator, which also serves as one of the electrical contacts to the piezo. The crystal sample is maintained at the sample bias voltage $V_{bias}$, while the other end of the piezo actuator is maintained at the voltage $V_{bias}+V_{strain}$ via a low-noise floating voltage supply. STM imaging is performed on an in-situ cleaved crystal, where the tunneling current is measured from the tip as usual. The strain is independently measured using a resistive strain gauge as well by interferometry. E-STS as implemented is broadly applicable to crystal as well as film samples.

We perform E-STS on the iron-pnictide parent compound NaFeAs[52]. This material has two known phase transitions – a transition from the high temperature tetragonal state to an orthorhombic state at $T_S$=52K, and a transition to a long-range ordered spin-density wave (SDW) phase at $T_{SDW}$=41K [53]. Electronic nematicity can be visualized in STS images as anisotropic patterns in the vicinity of defects that reduce the rotational symmetry from four-fold ($C_4$) to two-fold ($C_2$)[19, 20, 25]. In NaFeAs[19], and also in the related materials $CaFe_2As_2$[16] and FeSe[20], the reduction from $C_4$ to $C_2$ symmetry is primarily seen in electronic states close to the Fermi level suggesting that it originates from a low-energy instability of the electron gas.

We first discuss E-STS measurements performed on NaFeAs at low temperature (T=6K). In this regime the material has both long-range magnetic (stripe SDW) and structural (orthorhombic) order. Global constraints on the strain state, presumably arising from the details of sample mounting, mean that the orthorhombic order is not uniform across the sample. Instead, micron sized domains with near atomically sharp domain boundaries appear. The domain size presumably is set by the competition of domain wall and strain energies, and boundaries between domains have been previously visualized in STS imaging measurements[16, 19, 20].

In panels a-e of Fig. 1 we show images of the exact same region of our NaFeAs sample taken at a temperature of 6K for different values of the external strain, as indicated by the piezo voltage $V_{strain}$ (a larger set of images is shown as a movie in the supporting information, section SII). We tune the strain by changing the voltage applied to the strain piezo from +250V (maximal compression, panel a) to -250V (maximum tension, panel e), and measure the response of the material via STM. Imaging is performed at identical conditions at the same energy for each value of applied strain. For each new value of strain, the crystal undergoes a translational shift under the STM tip, providing an independent, in-situ measurement of applied strain. To zero out this shift we re-center the tip at each value of strain. In Fig. 1a two sharp domain walls are seen separating domains in which the direction of the longer (a) axis changes by ~90 degrees. On either side of the domain boundary, LDOS patterns that have $C_2$ symmetry can be seen in the vicinity of defects. These patterns also rotate by ~90 degrees across the domain boundary in accord with the structural distortion. Comparison to Fig. 1b-e shows that the domain walls move as a function of applied strain so that as the magnitude of the compressive strain is decreased, the area of the domains with long axis aligned parallel to the strain direction increases and the area of the domains with long axis perpendicular to the strain direction decreases. For each value of strain, we can calculate the ratio of the domain areas of the two orientations. We plot this ratio as a function of strain in

Fig. 1g. This figure shows the presence of significant hysteresis as well as irregular motion as a function of strain, reminiscent of Barkhausen noise[54] in the motion of domain boundaries at the atomic scale.

We can directly quantify the magnitude of electronic $C_4$ symmetry breaking as a function of strain from Fig. 1a-e. To do this, we take advantage of the fact that individual defect signatures can be easily identified in these images. We proceed by cropping a region centered on each defect and averaging all these cropped images together, thus generating an average spectroscopic signature for each value of strain (see supplementary info section SIII for more details of this procedure). This average defect structure is then subtracted from the same image rotated by 90 degrees. Any intensity in the resultant subtracted image is due to breaking of $C_4$ symmetry in the local electronic structure, and we can sum up all the intensity in the subtracted image to obtain a measure of the uniaxial electronic anisotropy at each value of strain. Since the images at different values of strain are obtained under the same tunneling conditions, we can directly compare the magnitudes of the anisotropy obtained by this process at different strain values. The resultant anisotropy is plotted as a function of strain in figure 1h. We see that within experimental error, the magnitude of the anisotropy is independent of applied strain at the low temperature at which this measurement was performed.

The data in Fig. 1 demonstrate the power of the E-STS technique to reveal the interplay between strain and electronic anisotropy. For an Ising-nematic ($C_4$-$C_2$ symmetry breaking),[54] the ground state contains domains of differently oriented orders. In the presence of uniaxial strain, one orthorhombic domain is favored over the other[55]. Our measurements of domain wall motion and anisotropy strength visualized in Fig. 1 rule out several competing scenarios for how electronic anisotropy and domain walls could evolve under strain. For instance, a competing scenario to that seen in experiment would be one where the domain walls are pinned (by disorder or collectively) while the magnitude of the electronic anisotropy is modified within each domain as a function of strain. Instead, our experiments indicate that nematic domains behave similarly to Ising ferromagnets, where an applied magnetic field changes the domain structure but does not affect the saturation magnitude of the magnetization within a domain.

We now turn to E-STS measurements at 54K, above $T_S$. Figs. 2a and 2b compare the STS images for our most negative (Fig. 2a) and most positive (Fig. 2b) strain voltages. Both images are obtained at the same tunneling conditions. A strong anisotropy (identified by vertical yellow streaks) is visible in the +200V image (Fig. 2b) while in the -200V image (Fig. 2a) the anisotropy has completely disappeared from the entire field of view. This behavior is also confirmed in the Fourier transforms (FT) of the STS data presented in Fig 2c and 2d. The FT of the +200V data shows strong $C_4$ symmetry breaking, with a pronounced three-peak structure. As previously reported[16], the three-peak structure is a signature of Fermi surface reconstruction, whose presence at temperatures $T>T_{SDW}$ we interpret as evidence for large-amplitude SDW fluctuations [56, 57]. In contrast, the FT for the -200V data shows a strongly diminished intensity overall along with weak (if any) $C_4$ symmetry breaking.

To interpret the high temperature E-STS data we first observe that experimental samples experience built-in strain, arising from sample growth, from the process of incorporation into the device, and from differential thermal contraction when cooled to low temperature. To reach a zero strain situation, external strain must be applied to counteract the built-in strain[3]. The fact that the electronic anisotropy is nearly destroyed at $V_{strain}$ = -200 V indicates that at this voltage the built-in strain is cancelled by the externally applied strain. In general, we find that different samples require different applied voltages $V_{strain}$ to eliminate the anisotropy, indicating differing values of built-in strain. We have thus established that at temperatures $T>T_S$ the physics is strongly affected by local strain, which in our apparatus can be continuously dialed to zero.

At high temperature we observe neither the hysteresis nor the nematic domains (see below for the temperature evolution of domain walls, and supplementary section SVI) that were characteristic of the response in the long-range ordered state for T<$T_{SDW}$. In the low T state, strain only affects the relative areas of the different domains, but not the locally determined magnitude of the anisotropy. In sharp contrast, at high T above $T_S$ there is no evidence of domains and the local value of the anisotropy depends on the applied strain. The fact that for a certain value of applied strain the electronic anisotropy can be reduced to a nearly vanishing value at every point in a wide field of view is conclusive evidence that the electronic structure above $T_S$ does not exhibit long-range nematic order. This implies that the anisotropic data shown in Fig. 2 can be interpreted as a para-nematic response of the electronic structure to applied strain: the strength of the electronic anisotropy observed is dependent on the net strain applied to the field of view studied in the experiment.

Our observation that the magnitude of the anisotropy can be controlled by strain at high temperature is a key new insight that E-STS provides. We will show below that the intensity of the electronic anisotropy seen in STM is directly related to the amplitude of the nematic fluctuations. Thus our data show that the amplitude of the nematic fluctuations themselves are set by the strain applied to the system. Such a strong nonlinear coupling between the structure and electronic nematicity has not previously been anticipated, and indicates the importance of properly accounting for the structural degrees of freedom in any description of the electronic properties of the pnictides.

To present the anisotropy in a form that can be compared more directly to theory we subtracted the FT image from its rotation by 90 degrees (see supplementary material section SIII and SV for details) to create difference plots (Fig 2e and 2f). In a $C_4$ symmetric situation the result would be zero up to noise, and indeed for the most negative strain voltage (2e) little anisotropy is visible; however for the highly strained case (2f) a strong $C_4$ symmetry breaking is visible. We have modeled the STS data theoretically along the lines of our previous work[19] by computing the density of states modifications arising from unidirectional (stripe) SDW fluctuations at T>$T_{SDW}$ (see the supplementary material section SIV for details). The calculations depend on three parameters: the amplitude of the incoherent SDW fluctuations denoted $\Delta_{LRA}$, the correlation length $\xi$ and the size of the Fermi pockets. Comparisons of Figs 2g-h to 2e-f reveal a nice qualitative and quantitative agreement. Modelling the change from Fig. 2f to 2e (or Fig. 2h to 2g) requires a substantial change in $\Delta_{LRA}$, as a decrease in the correlation length is not enough. In other words the amplitude of the incoherent SDW fluctuations in the paramagnetic/para-nematic state is itself a strong function of strain.

Having established the role of strain in the electronic anisotropy observed in STS, we present a quantitative measure of the electronic anisotropy as a function of temperature. To do this, we track a constant area of the sample as a function of temperature while keeping the externally applied strain to zero. Shown in Fig. 3a-f are a sequence of STS images taken over the same area of the sample as a function of temperature starting from T=28K < $T_{SDW}$, through T=52K > $T_S$. These images are all taken under identical tunneling conditions, and each temperature is stabilized for approximately a day before the measurement is performed. The differential thermal expansion between the sample and substrate over this range of temperature is estimated to be <0.01% and can thus be neglected. The images show several interesting features. First, we note the presence of domain boundaries (visible as light stripes, marked with arrows for clarity) in Fig. 3a-e[16, 19, 20] (there is no domain wall visible in panel f). It is seen that as the temperature is raised, the position of the domain walls changes in such a way that the area of the minority domain decreases and the image contrast that defines the domain wall decreases. Exactly at $T_S$, domain walls completely disappear from the image (figure 3f). To confirm that domains disappear in a much larger field of view than that presented in figure 3, we have tracked ~ 500 nm x 500 nm areas of the sample across the structural transition temperature and have confirmed that the domain walls disappear at $T_S$ (see supplementary section VII for these images). Further, we have extensively scanned over 80 $\mu m^2$ of sample area both below and above $T_S$ and have never observed domain walls above $T_S$. Our temperature dependent data for domain walls together with the strain dependence of the anisotropy provide definitive microscopic evidence that the true nematic transition is at the structural transition temperature $T_S$ and that there is a strong nematic susceptibility above $T_S$. While a similar conclusion has previously been reached by transport measurements and has been conjectured by some of us, the new data provide microscopic and spectroscopic evidence that rules out scenarios where a true nematic transition occurs above $T_S$.

We next use the temperature-dependent dataset in Fig. 3 to quantify the anisotropy seen in STM as a function of temperature. For each temperature, we determine the magnitude of $C_4$ symmetry breaking as in the analysis of Fig. 1 (see supplementary information section SIII) for the domain that survives across $T_S$. We plot the resultant magnitude of the anisotropy as a function of temperature as the red dots in Fig. 3g. In NaFeAs the spin density wave transition temperature $T_{SDW}$ (41K) is clearly separated from the structural transition temperature $T_S$ (52K) allowing us to distinguish the effects of the different orderings on the nematic order parameter. The most significant feature of this plot is the presence of a clear kink in the data just below the bulk $T_{SDW}$. Several measures have been taken to minimize statistical and systematic sources of error in this plot. While the number of points on the curve is limited by the total data acquisition time (several months), each data point represents an average over several hundred defects in the field of view and thus has virtually no statistical error. Independent data sets have also been obtained during heating and cooling experiments with identical results, and additional data that shows the same result for the orthogonally oriented domain is shown in supplementary section VIII. The number of temperature data points we are able to obtain is constrained by the probability of tip changes at the elevated temperatures, and we optimize the experimental run time to keep the tip and sample conditions identical in the important temperature range of 25-55 K, eliminating matrix element changes as a source of error. Thus, the observed kink is a true feature of the data set. The observed sharp decrease near $T_{SDW}$ rather than $T_S$ is direct evidence that the electronic nematicity observed in the electronic structure is primarily driven by spin fluctuations in this iron-based compound. We note that a close examination of the experimentally determined anisotropy parameter shows that the kink occurs a few Kelvin below the bulk $T_{SDW}$. Potential reasons for this include a slightly different surface $T_{SDW}$ and disorder-induced inhomogeneity in the locally measured $T_{SDW}$.

To understand the consequences of these measurements, we have developed a theoretical model to compute the QPI signal resulting from a Fermi surface reconstruction arising from either long-ranged stripe SDW order (at $T<T_{SDW}$) or stripe SDW fluctuations (at $T>T_{SDW}$) or a combination of the two. The model (see the supplementary material section SIV for details) involves a gap parameter $\Delta_{SDW}$ parametrizing the amplitude of the fully coherent SDW order that sets in below $T_{SDW}$, a gap parameter $\Delta_{LRA}$ parametrizing incoherent SDW amplitude fluctuations (nematic fluctuations) at $T>T_{SDW}$ and the correlation length $\xi$ introduced above. The results, shown in Fig 3g, confirm that a single model based on unidirectional SDW order and fluctuations can fully account for the experimentally observed anisotropy across different temperatures. To pinpoint whether the change observed at $T_{SDW}$ arises from the onset of coherent long ranged order or from a change in the amplitude of the (fluctuating plus coherent) gap we show two alternative calculations. In the first scenario, which highlights the impact of coherence factors, the magnitude of the total gap $\Delta^2=\Delta^2_{SDW}+\Delta^2_{LRA}$ remains constant below $T_{SDW}$ but there is a transfer of incoherent spectral weight to coherent spectral weight as temperature is lowered, i.e. $\Delta_{SDW}$ increases at the same rate as $\Delta_{LRA}$ decreases. Then, the only change below $T_{SDW}$ is the appearance of coherence factors (via the anomalous, momentum off-diagonal Green's function). In the second scenario, which highlights the effects of fluctuations, the fluctuating gap $\Delta^2$ increases as temperature is lowered due to an increasing $\Delta_{SDW}$ and a constant $\Delta_{LRA}$. The modeling clearly demonstrates that the main cause of the rapid increase in the anisotropy parameter below $T_{SDW}$ is an increase in the magnitude of the total fluctuating gap $\Delta^2=\Delta^2_{SDW}+\Delta^2_{LRA}$, whereas the coherence factor effects arising from long-range SDW order play a minor role. We have also modeled the strain dependence at $T>T_{SDW}$ by varying the fluctuating gap and the correlation length (see Supplementary Material section IV for details).

We now study the entire doping and temperature phase diagram of the system $NaFe_{1-x}Co_xAs$. Fig. 4a shows a real space STS scan of $NaFe_{(1-x)}Co_{(x<0.01)}As$ at 6K, well within the long-ranged magnetically ordered phase ($T<T_{SDW}$), showing several domain boundaries like those seen in the parent compound. Domain boundaries are observed only for dopings x < 0.02 in agreement with previous [53] specific heat and resistivity measurements that establish bulk long-range structural order. At higher doping, it becomes difficult to directly visualize the electronic anisotropy in real space due to the large number of dopants. However, we can study the prevalence of anisotropy in the images by studying their Fourier Transforms[19, 25] and observing their $C_2$ or $C_4$ symmetry. For regions of the phase diagram that display domains, measurements are conducted within a single domain. Figure 4c-i shows the evolution of the FT of STS images near the Fermi energy (cropped to one-half of the first Brillouin Zone) for different doping concentrations and temperatures. While we clearly observe $C_2$ symmetry in most of the phase diagram, we have never observed domains or domain boundaries at dopings and temperatures outside the regime of long ranged SDW order (i.e. Fig. 4e-g). This indicates the disappearance of long-range order but the persistence of nematic fluctuations as doping is increased beyond the critical doping for the SDW order, similar to the phenomena we observed in the parent compound as temperature is increased above $T_S$. A close examination of the FTs in Fig 4c-i shows that while the details of the patterns continuously evolve with doping, the wavevectors at which anisotropy is observed are fairly similar in magnitude across the phase diagram. ARPES measurements[58, 59] of the doping-dependent bandstructure in $NaFe_{(1-x)}Co_{(x)}As$ show fairly small changes in the dispersions of the bands and the chemical potential from the parent compound to the overdoped side of the phase diagram. Our work on the parent compound described above shows that in order to get QPI features at wavevectors similar to experiment, it is essential to include band folding due to spin density wave order or fluctuations. This indicates that the anisotropy observed in our STS data comes primarily from spin order and fluctuations even for doped samples.

Our doping dependent measurements show that electronic anisotropy exists both inside (Figure 4e-g) and outside (Figure 4h) the superconducting dome. We do see changes in the strength and wavevectors of the anisotropy across $T_C$ (for example, comparing figure 4g and 4h), and detailed future measurements that track the same area of the sample across $T_C$ can help address the interplay between the superconducting gap and nematic fluctuations. For the highly overdoped, non-superconducting sample we see $C_4$ symmetry in the FT, indicating the absence of any long-range nematic order or nematic fluctuations (Fig. 4i). A detailed exploration of the phase diagram in this region can help in understanding the interplay between nematicity and superconductivity in these compounds[1, 10]. The co-existence of these two phenomena and how each of them responds to strain provides an interesting dynamic for further study, especially given the evidence for a nematic quantum critical point in the superconducting dome in some families of the pnictides[60].

In summary, we have introduced a new E-STS technique that enables investigation of the strain dependence of electronic properties with real-space atomic resolution. With this method we are able to experimentally distinguish between long-range electronic plus lattice nematic order (marked by the presence of domain walls and hysteresis under switching of strain) and purely electronic asymmetry due to fluctuations in the presence of strain (marked by $C_2$ symmetry in the response to defects). We showed that the dominant source of electronic nematic response is antiferromagnetic (stripe) spin fluctuations and found that the amplitude of the slow stripe fluctuations in the paramagnetic phase depends strongly on the magnitude of the actual strain felt by the electrons. We note that the energy scale associated with nematicity observed in STM remains several tens of meV even on the overdoped side, indicating that there is not a direct proportionality between superconducting $T_C$ and the energy scale associated with nematicity. At the same time, our STM measurements show a clear decrease in the intensity of the nematicity as a function of increasing doping, pointing to an interesting intertwining with superconducting order. By extending our E-STS measurements to the optimal and overdoped range iron-based compounds, we can gain insight into the relationship between nematicity order, fluctuations and superconductivity.

This work is supported by the National Science Foundation via grant DMR- 1610110 (E.F.A., A.N.). Salary and materials support is also provided by the Air Force Office of Scientific Research via grant number FA9550-16-1-0601 (E.R., A.N.P.). R.M.F. and X.W. are supported by the Office of Basic Energy Sciences, U.S. Department of Energy, under award DE-SC001233 6. X.W. is also supported by a Doctoral Dissertation Fellowship in University of Minnesota. and A.J.M. by the National Science Foundation under grant DMR-1308236. Work at IOPCAS is supported by NSF & MOST of China through Research Projects, as well as by the CAS External Cooperation Program of BIC (112111KYS820150017).

**Figure Captions:**

**Figure 1: Elasto-STM in the nematic ordered state**
(a-e) STM images of NaFeAs at T=6K at various values of applied strain, obtained on the exact same area of the sample (white bar represents 10nm). Nematic order can be visualized as yellow, unidirectional streaks in the image. At this temperature, nematic domains exist, with the domain boundaries showing up as dark lines in the image. As the applied strain (proportional to the voltage $V_{strain}$ shown in the bottom of the panels) is changed in the sample, the domain walls move and reduce the area of one of the domains at the expense of the other. Imaging conditions V=+10 mV, I=100 pA. (f) Apparatus used to strain samples while STM is being performed on them. (g) Plot of the ratio of the size of each of the domains in the field of view in images (a-e) as a function of applied strain. It is clearly seen that the motion of the domains exhibit hysteresis and domain pinning. (h) Intensity of the nematic anisotropy parameter as a function of the applied strain. Within experimental error the intensity is independent of strain.

**Figure 2: Elasto-STM in the para-nematic state**
(a-b) STM spectroscopy images of NaFeAs at T=54K > $T_s$ over the same area of the sample at two values of applied strain, -200 V and +200 V respectively (white bar represents 20nm). Spectroscopy is performed at V=+20 mV and I=100 pA. It is seen that the rotational anisotropy is small at -200 V and largest at +200 V. (c-d) Absolute value of smoothed Fourier transform of the STS images with applied strain, -200V and +200V. While (c) is nearly $C_4$ symmetric, (d) shows clear anisotropy. The overall magnitude of the response is also clearly smaller in (c). (e-f) Show difference plots obtained by rotating the FT in (c-d) by 90 degrees and subtracting them from the un-rotated FT (c-d), respectively. Non-zero values indicate $C_4$ symmetry breaking. (g-h) Difference plots obtained from theoretical modeling of the QPI arising from unidirectional SDW fluctuations, with (g) having smaller incoherent fluctuation amplitude than (h).

**Figure 3: Nematic domains and the driving force behind electronic nematicity**
(a-f) STM spectroscopy images of NaFeAs at various temperatures over the same area of the sample (white bar represents 20nm). Spectroscopy is performed at V=+10 mV and I=100 pA. At low temperature, nematic domains are seen in the sample as straight lines, and the orientation of the nematic order rotates by ~ 90 degrees across the domain boundary as is also seen in the Fourier images of each side of the domain (inset to (a)). As the temperature is raised, the domains move and eventually vanish at the structural phase transition $T_S$=52 K. (g) Nematic anisotropy measured as a function of temperature (red dots, note that the data are averaged over defects as described in supplementary material section SIII) and calculated from theory as described in text (solid curves). It is seen that a sharp kink exists at the magnetic transition temperature $T_{SDW}$=41 K, and most of the intensity in the nematic signal picks up only below the magnetic transition temperature. The key ingredient in the optimum theoretical fit (blue line) is the increase in the total SDW gap (i.e. both coherent and incoherent contributions) below $T_{SDW}$, as opposed to the appearance of coherent factors only (green curve). Note that removing the coherence factors (orange curve) barely changes the behavior of the blue curve.

**Figure 4: Doping dependence of nematic order and fluctuations in NaFe$_{1-x}$Co$_x$As**
(a) STM spectroscopy images of NaFe$_{1.99}$Co$_{0.01}$As showing the presence of domain structures (white bar represents 40nm). A superconducting gap (not shown) is also observed at this temperature. (b) Phase diagram of observed nematic domains and nematic anisotropy from STM measurements, superposed with bulk measurements of the phase diagram. We note that domains are only observed in STM where bulk orthorhombic order is known to exist (orange shaded region). In the striped region, domains and superconducting gaps are observed to coexist in space by STM. Nematic anisotropy is observed across the phase diagram and only disappears for samples that are not superconducting. (c-i) Nematic anisotropy as a function of doping in Fourier space. Shown are a sequence of Fourier transforms of spectroscopy images obtained on $NaFe_{1-x}Co_xAs$ for various values of x. All images are obtained at V=+10 mV. It is seen that the shape of the nematic structure in Fourier space evolves with doping, but the anisotropy itself persists across the superconducting dome and only disappears for x=0.12 which is beyond the superconducting dome.

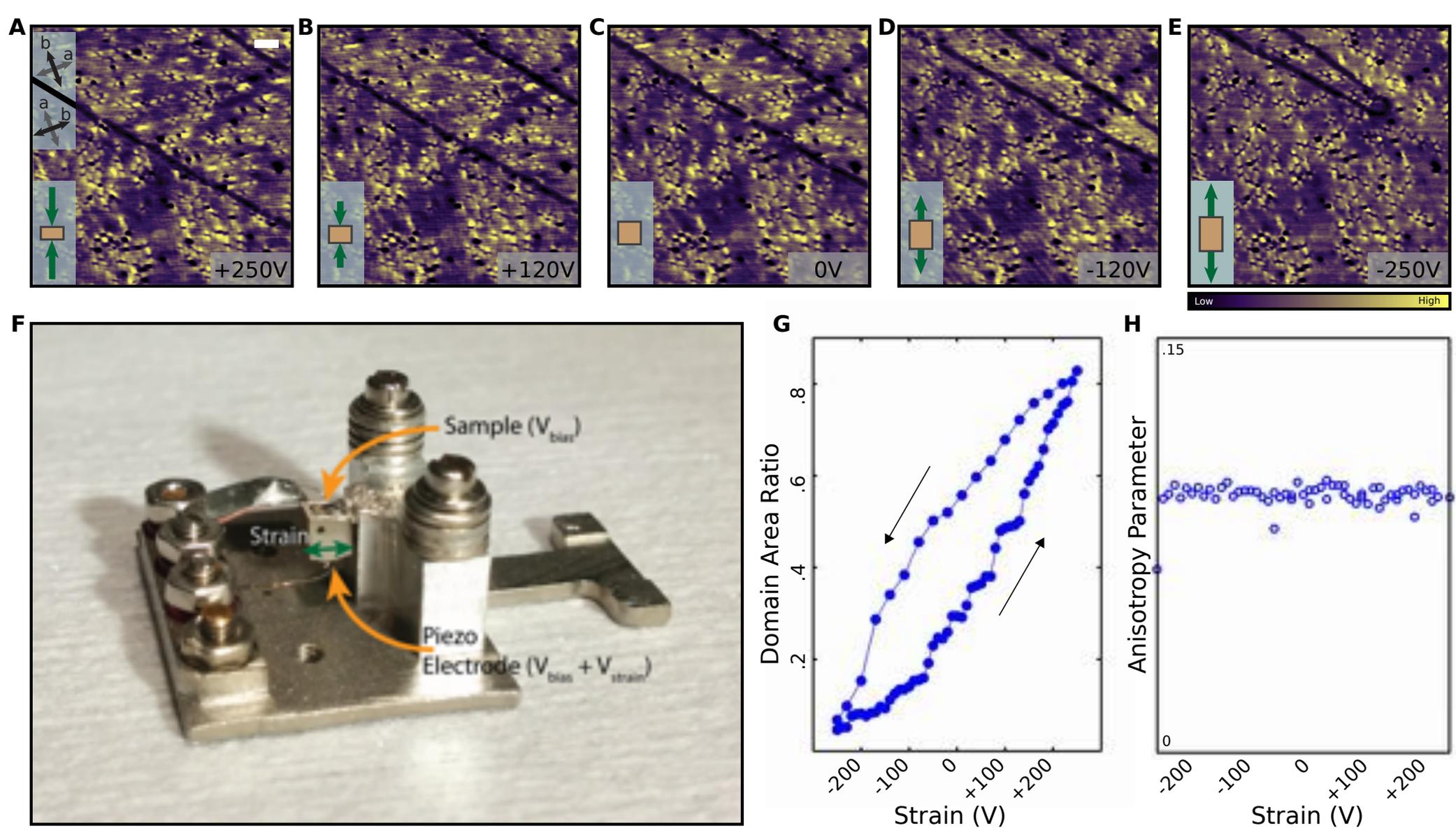

Figure 1

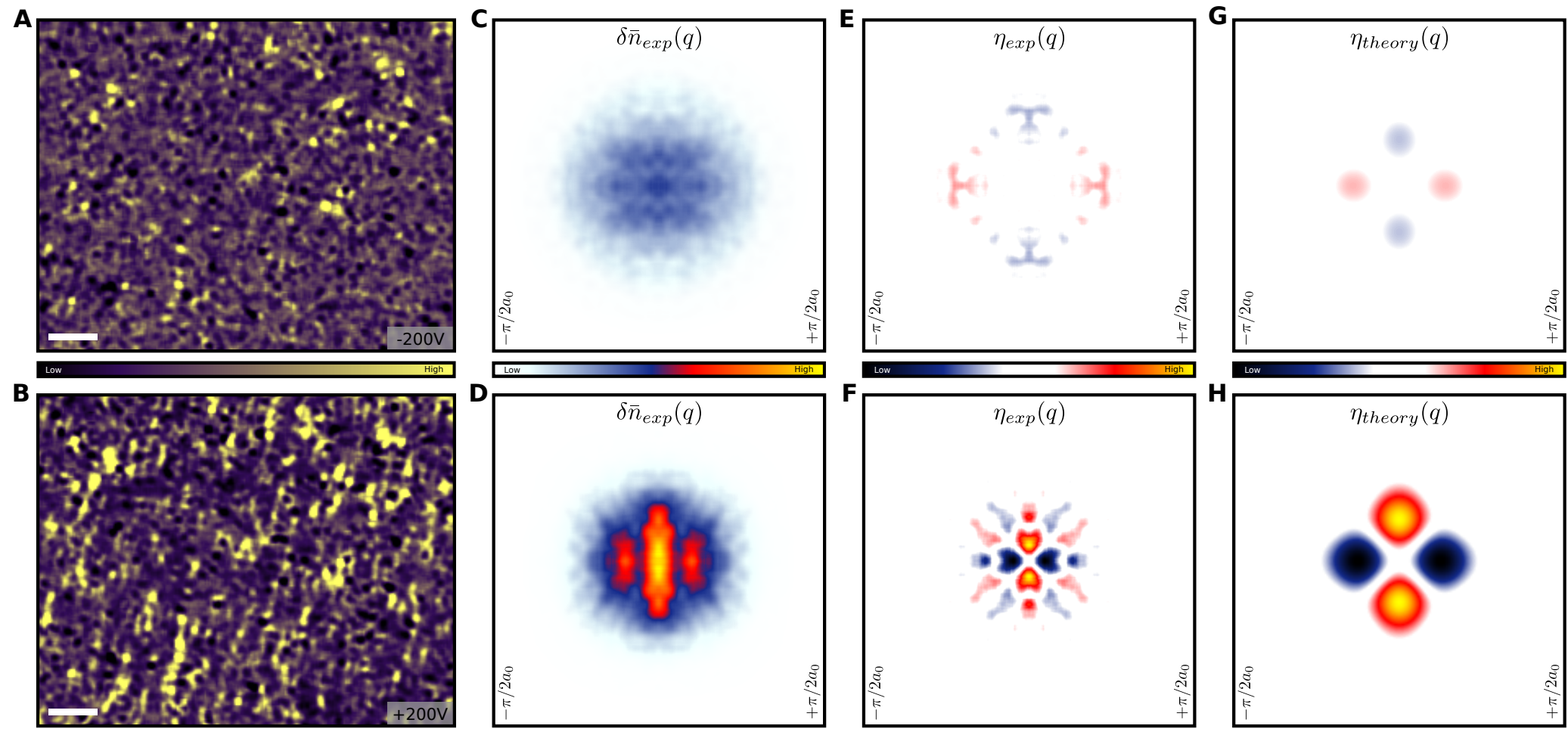

Figure 2

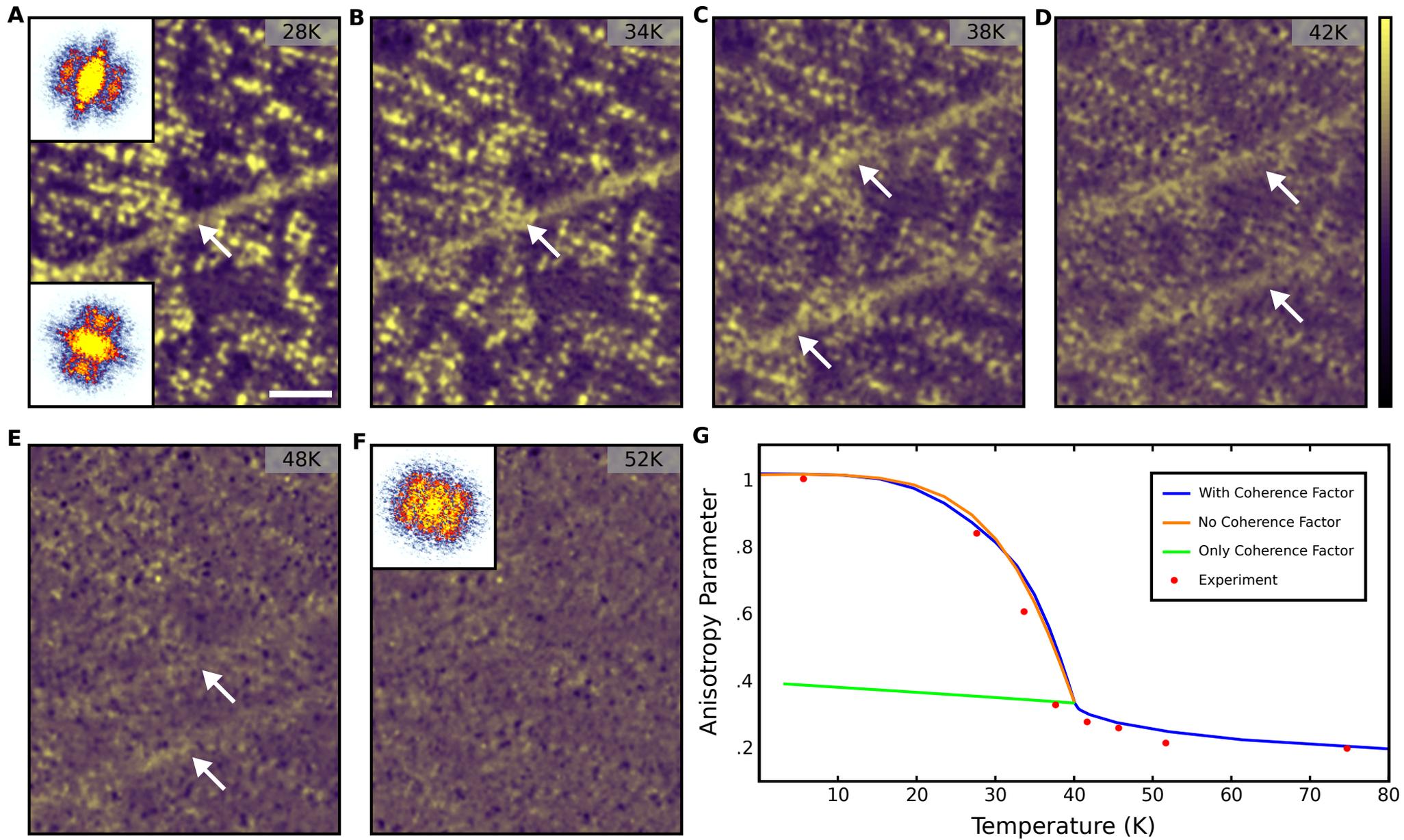

Figure 3

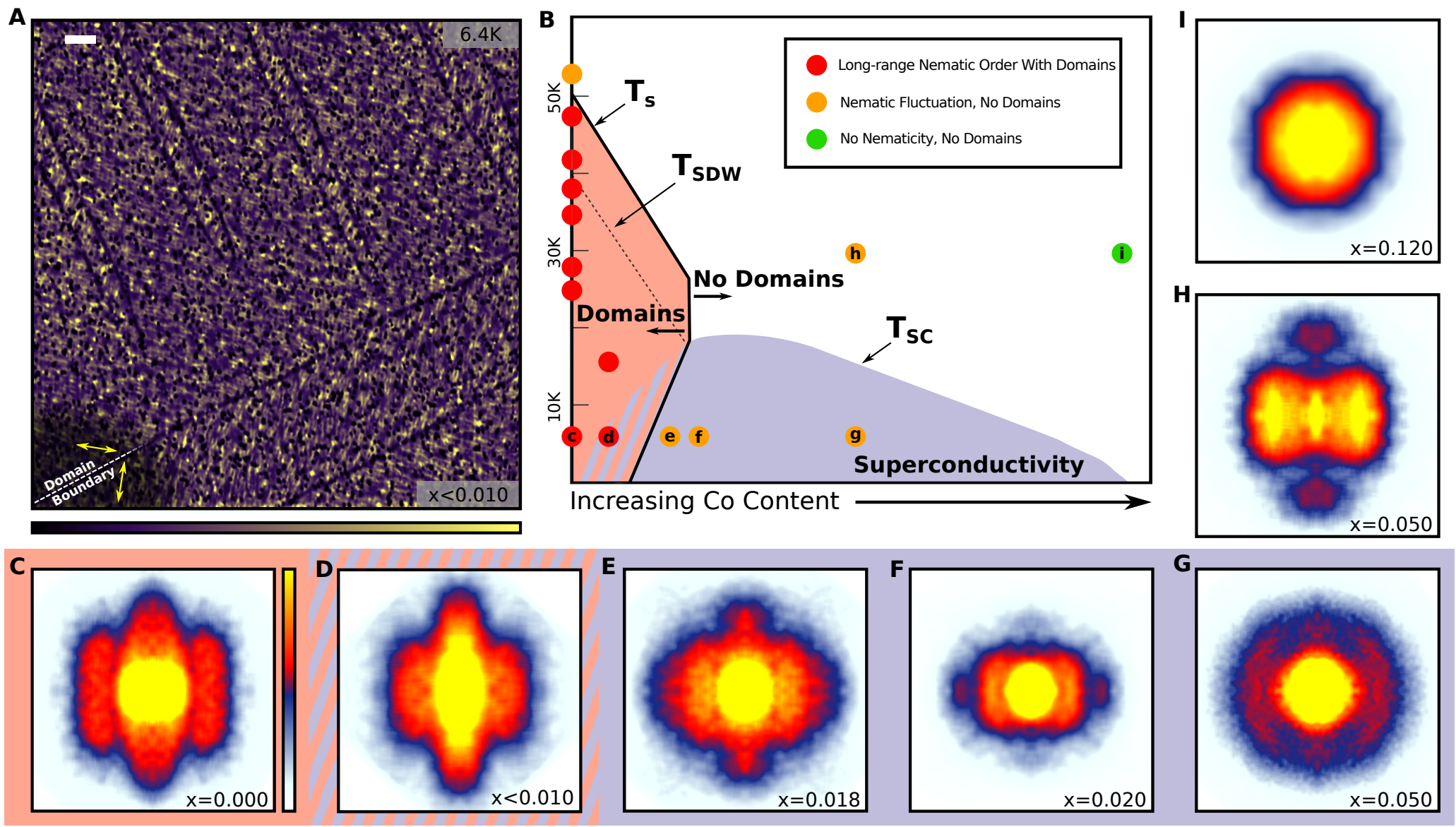

Figure 4

# Supplementary Material

Authors: Erick F. Andrade[1], Ayelet D. Notis[1], Ethan P. Rosenthal[1], Xiaoyu Wang[2], Lingyi Xing[3], Xiancheng Wang[3], Changqing Jin[3], Rafael M. Fernandes[2], Andy J. Millis[1], Abhay N. Pasupathy[1]

Affiliations: [1]Department of Physics, Columbia University, New York, NY 10027, USA. [2]School of Physics and Astronomy, University of Minnesota, Minneapolis, MN 55455, USA. [3]Beijing National Laboratory for Condensed Matter Physics, Institute of Physics, Chinese Academy of Sciences, Beijing 100190, China.

## I. STM Data acquisition and image processing

STM $dI/dV$ maps were obtained using standard lock-in techniques with oscillation frequencies between 1.5 and 1.8 kHz. To generate Fourier space images shown in the main text, the following procedures are performed:
(a) The real space image is affine transformed to remove any scan drift in the image
(b) A two-dimensional magnitude FFT is computed from the transformed real space image in MATLAB which assumes periodic boundaries.
(c) The FFT image in (b) contains a spike at $q=0$ due to the nonzero average value of the $dI/dV$ over the entire image. Since this value plays no role in our analysis of QPI or anisotropy, it is removed by setting the $q=0$ pixel to the average of its nearest neighbors.
(d) The FT in (c) is rotated to make the $q_x$ axis horizontal
(e) The FT in (d) suffers from a noise that arises from the random positions of defects that give rise to QPI signals. Taking the magnitude of the FT provides some improvement, however, significant amplitude noise exists in the FT even after taking the magnitude. Two operations are performed to improve the signal to noise. First the FT is mirror symmetrized along the principal axes. If $\delta n(\boldsymbol{q},\omega)$ represents the magnitude of the FT after (d), the mirror symmetrization generates the FT
$$\delta\tilde{n}(\boldsymbol{q},\omega) = \left[\delta n(\boldsymbol{q},\omega) + \delta n(R_{q_x}\boldsymbol{q},\omega) + \delta n(R_{q_y}\boldsymbol{q},\omega) + \delta n(R_{q_x}R_{q_y}\boldsymbol{q},\omega)\right]/4$$
with $R_a$ denoting a mirror reflection along axis $a$. Second, a Gaussian filter is applied to generate the final FT: $\delta\bar{n}(\boldsymbol{q}) = \int \delta\tilde{n}(\boldsymbol{q}')F(\boldsymbol{q}-\boldsymbol{q}')d\boldsymbol{q}'$ where $F(\boldsymbol{q}) = e^{-q^2/2\sigma^2}$, $\sigma$ is .7% of the 1 Fe first Brillion zone, and the filter size is 2% of the 1 Fe first Brillion zone.

## II. Low Temperature E-STM Movie

We acquired a total of 68 LDOS images at a bias of +10 mV while varying the strain from -250V to +250V and back down to -250V. We manually line up the images with each other using the defect positions (which do not change with strain) and compile them to produce a movie.

## III. Quantifying Anisotropy in STS images

We outline below the procedures used to calculate the C₄ anisotropy in STS images in Fig. 1-3 of the main text. In STS scans where the defect density is low (such as those in Fig. 1 and 3 of the main text) and the real space positions of the defects can be clearly identified, it is easy to directly identify the real space QPI signal corresponding to a single defect. To do this, we first identify manually the positions of each one of the defects in real space. We crop a small <10nm square region around each identified defect and average all of the cropped regions together to produce an average real space QPI signal associated with a single defect. To quantify the anisotropy in this average image, we rotate the image by 90 degrees (**r**→**r̃**) and subtract it from itself, to generate a real space "difference plot". Any non-zero value in the difference plot comes from C₄ symmetry breaking in the original image. We then obtain the anisotropy parameter **η** by summing the absolute value of each pixel in the difference plot and normalizing by the sum of absolute values in the original image before rotation and subtraction. Mathematically,

$$\eta(\omega) = \frac{\sum_{\mathbf{r}}' |\delta n(\mathbf{r},\omega) - \delta n(\mathbf{\tilde{r}},\omega)|}{\sum_{\mathbf{r}}' |\delta n(\mathbf{r},\omega) + \delta n(\mathbf{\tilde{r}},\omega)|}$$

where **r̃** is the 90 degree rotation of **r** about the $z$ direction and the prime indicates the sum over the smaller region of the average defect.

In the presence of large numbers of defects where individual defects cannot be identified with certainty, as well as cases where the QPI signal is weak relative to other spectroscopic features present in the sample, we cannot directly identify the anisotropic signature due to a single defect experimentally without additional modeling. This is the case in NaFeAs at high temperature in the presence of strain (Fig. 2 of the main text). In this situation, we work in Fourier space and consider the FT of an entire STS image $\delta \bar{n}(\mathbf{q},\omega)$. To calculate the anisotropy in such an image, we rotate the image by 90 degrees (**q**→**q̃**) and subtract it from itself to generate a Fourier space difference plot. We discuss the relationship between these experimentally calculated measures of anisotropy and theory in section V below.

**IV. Theoretical Overview**

The relevant experimental quantity is the derivative with respect to bias voltage $V_{bias}$ of the sample-tip tunneling current $I$ ($dI/dV$), measured as a function of position **r** on the surface of the sample. In an ideal sample the measured quantity would have the full translational and rotational symmetry of the lattice, but in the presence of defects at positions $\mathbf{R}_i$ the measured quantity varies with position in a manner which is believed to be proportional to the defect-induced change $\delta n(\mathbf{r},\omega)$ in the local electronic density of states (LDOS) at position **r** and energy $\omega = eV_{bias}$. The LDOS in turn is related to the defect scattering potentials $V_i$ by an electronic susceptibility $\chi$ that encodes information about the electronic physics and is discussed in more detail below:

$$\delta n(\mathbf{r},\omega) = \sum_i \chi(\mathbf{r} - \mathbf{R}_i, \omega) V(\mathbf{R}_i) \qquad \textbf{(S1)}$$

Here we have assumed (as is the case in the experiment) that the resolution is on the scale of the unit cell size or greater so we may take the susceptibility to be translation-invariant and neglect local field corrections. For simplicity we also assume, following standard practice in the STS field, that each defect is point-like and gives rise to the same scattering potential and that the scattering is weak enough that a linear-response ansatz for the electronic response suffices.

The physical information is carried by the response function $\chi$, which describes the electronic standing waves created around each defect. The interference of the standing waves from different randomly positioned defects creates complicated patterns. We have found by modeling situations with different densities that if the pattern associated with an individual defect cannot be isolated, the procedure of smoothing as described in section I above and then computing the difference of the image and its 90 degree rotation provides the best way to extract information about $\chi$.

*Model:* We use a four-band model of the pnictide Fermi surface with two zone-center hole-like Fermi pockets, labeled by $\gamma_1$ and $\gamma_2$, and two elliptical electron pockets, labeled by $X$ and $Y$. In the Brillouin zone appropriate to the single-Fe unit cell, the two electron pockets are centered at the $\mathbf{Q}_x = (\pi, 0)$ and $\mathbf{Q}_y = (0, \pi)$ points.

The band dispersions of the two hole pockets are given in terms of a function $h(\alpha_\gamma) = -\mu_\gamma + b\left(2\alpha_\gamma(1 - \cos k_x) + \frac{2}{\alpha_\gamma}(1 - \cos k_y)\right)$ as

$$\varepsilon_{\gamma_1 \mathbf{k}} = \frac{1}{2}\left(h(\alpha_\gamma) + h(\frac{1}{\alpha_\gamma})\right) + \frac{1}{2}\sqrt{\left(h(\alpha_\gamma) - h(\frac{1}{\alpha_\gamma})\right)^2 + W^2}$$

$$\varepsilon_{\gamma_2 \mathbf{k}} = \frac{1}{2}\left(h(\alpha_\gamma) + h(\frac{1}{\alpha_\gamma})\right) - \frac{1}{2}\sqrt{\left(h(\alpha_\gamma) - h(\frac{1}{\alpha_\gamma})\right)^2 + W^2}$$

while the band dispersions of the electron pockets are

$$\varepsilon_{X\mathbf{k}+\mathbf{Q}_x} = -\mu_e + b\left(2\alpha_e(1 - \cos k_x) + \frac{2}{\alpha_e}(1 - \cos k_y)\right)$$

$$\varepsilon_{Y\mathbf{k}+\mathbf{Q}_y} = -\mu_e + b\left(\frac{2}{\alpha_e}(1 - \cos k_x) + 2\alpha_e(1 - \cos k_y)\right)$$

The parameters $b = 4, \mu_\gamma = -0.17, \alpha_\gamma = -2, W = 0.12, \mu_e = 0.32$, and $\alpha_e = 0.4$ are chosen so that the resulting Fermi surface resembles the Fermi surface of NaFeAs measured by ARPES [3]. The Fermi surface is shown in Fig. S1. Here, all energy scales are measured in units of $\varepsilon_0 \approx 1/3$ eV, such that the bottom of the electron band is about 100 meV below the Fermi level.

To compute the effect of SDW order and fluctuations on this Fermi surface we follow Refs. [1,2]. We allow for the possibility of long-range stripe-like spin density wave order with a single ordering vector. For definiteness we choose the ordering wave-vector to be $\mathbf{Q}_x = (\pi, 0)$. SDW order is characterized by an order parameter $\langle \Delta(\mathbf{r}) \rangle = \Delta_{SDW}$. A non-zero $\Delta_{SDW}$ couples the wave-vector **k** to **k+Q**$_x$, in particular mixing the $X$ electron band to the two hole bands (for simplicity we include only the coupling to the band $\gamma_1$ with the larger Fermi surface) and opening a gap, thereby changing the dispersion. We also allow for the possibility that long-range order is destroyed by phase fluctuations, $\langle \Delta(\mathbf{r})\rangle \to 0$, so that there is no coherent coupling between **k** and **k+Q_x** while fluctuations in the amplitude $\langle \Delta(\mathbf{r})^2\rangle \equiv \Delta_{LRA}^2$ remain non-vanishing, so that a "pseudogap" is opened. We represent this situation mathematically via an 8x8 matrix electron propagator $\mathcal{G}_\sigma$ including both **k** and **k+Q_x** terms as

$$\mathcal{G}_\sigma = \begin{pmatrix} g_{\gamma_1} & 0 & 0 & 0 & 0 & \sigma f & 0 & 0 \\ 0 & \tilde{g}_{\gamma_1} & 0 & 0 & \sigma\tilde{f} & 0 & 0 & 0 \\ 0 & 0 & g_{\gamma_2} & 0 & 0 & 0 & 0 & 0 \\ 0 & 0 & 0 & \tilde{g}_{\gamma_2} & 0 & 0 & 0 & 0 \\ 0 & \sigma\tilde{f} & 0 & 0 & g_X & 0 & 0 & 0 \\ \sigma f & 0 & 0 & 0 & 0 & \tilde{g}_X & 0 & 0 \\ 0 & 0 & 0 & 0 & 0 & 0 & g_Y & 0 \\ 0 & 0 & 0 & 0 & 0 & 0 & 0 & \tilde{g}_Y \end{pmatrix}$$

where the normal (g) and anomalous (f) parts of $\mathcal{G}_\sigma$ are:

$$g_{\gamma_1}(\mathbf{k},\omega) = \left(\omega - \varepsilon_{\gamma_1\mathbf{k}} - \frac{\Delta_{SDW}^2 + \Delta_{LRA}^2}{\omega - \tilde{\varepsilon}_{X\mathbf{k}} + i\xi^{-1}}\right)^{-1}$$

$$\tilde{g}_X(\mathbf{k},\omega) = \left(\omega - \tilde{\varepsilon}_{X\mathbf{k}} - \frac{\Delta_{SDW}^2 + \Delta_{LRA}^2}{\omega - \varepsilon_{\gamma_1\mathbf{k}} + i\xi^{-1}}\right)^{-1}$$

$$g_Y(\mathbf{k},\omega) = (\omega - \varepsilon_{Y\mathbf{k}})^{-1}$$

$$g_{\gamma_2}(\mathbf{k},\omega) = (\omega - \varepsilon_{\gamma_2\mathbf{k}})^{-1}$$

$$f(\mathbf{k},\omega) = \frac{\Delta_{SDW}}{(\omega - \varepsilon_{\gamma_1\mathbf{k}})(\omega - \tilde{\varepsilon}_{X\mathbf{k}}) - (\Delta_{SDW}^2 + \Delta_{LRA}^2)}$$

$\sigma$ denotes the electron spin and a tilde denotes the same function evaluated at **k+Q_x**.

Following Ref. [2], we have also included a phenomenological broadening parameter $\xi$ measured in units of the lattice spacing $a$ and related to the correlation length of the phase fluctuations.

The extra terms in $g_{X,\gamma_1}$ express the effect of coherent and incoherent spin fluctuations in opening up a gap, while $f$ expresses the effect of coherent backscattering associated with long ranged order. We distinguish the fully normal phase ($\Delta_{SDW} = \Delta_{LRA} = 0$), the fluctuating nematic phase ($\Delta_{SDW} = 0, \Delta_{LRA} \neq 0$) and the ordered phase ($\Delta_{SDW} \neq 0, \Delta_{LRA} \neq 0$).

*QPI calculation, real space:* we now use standard formulas to compute the change in density of states, $\delta n(\mathbf{r},\omega)$, due to a non-magnetic impurity located at the origin. In the first Born approximation we have $\delta n(\mathbf{r},\omega) = Tr[\mathbf{M}\mathbf{G}(\mathbf{r},\omega + i\delta)\mathbf{V}\mathbf{G}(-\mathbf{r},\omega + i\delta)]$ with **G(r)** the Fourier transform of the G defined above, **M** the square of the matrix element linking the STM tip to the band states, and **V** the impurity scattering (all bold faced quantities are 8x8 matrices in the reduced zone defined above). We make the simplifying assumptions that the impurity scattering potential and STM matrix elements are momentum and band independent (connecting all

momenta to all momenta and all bands to all bands, with equal amplitudes). Carrying out the sum one finds

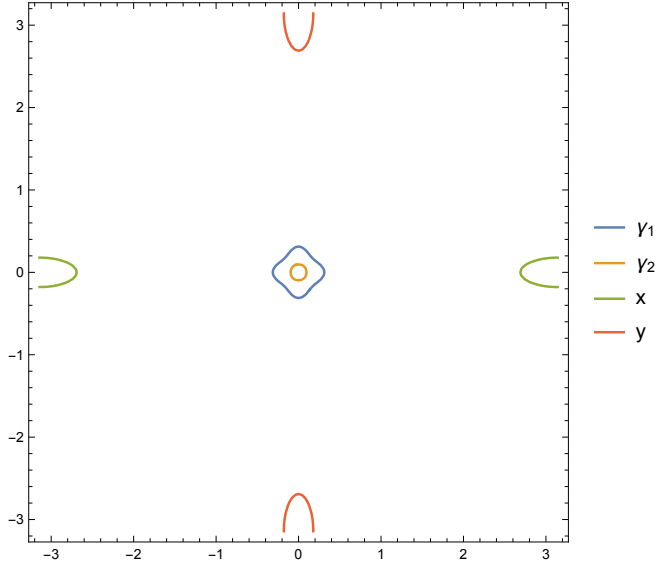

Figure S1: Fermi surface for the four-band model, paramagnetic phase.

$$\delta n(\mathbf{r}, \omega) = -\frac{VM}{\pi} \lim_{\delta \to 0_+} (\sum_\sigma \left[\sum_{ij} \mathcal{G}_{ij\sigma}(\mathbf{r}, \omega + i\delta)\right]\left[\sum_{kl} \mathcal{G}_{kl\sigma}(-\mathbf{r}, \omega + i\delta)\right] - \delta \leftrightarrow -\delta)$$

where the Roman indices denote elements of the $\mathcal{G}$ matrix defined above. Carrying out the sum over elements and spin degrees of freedom we find (the 2 is from the spin sum)

$$\delta n(\mathbf{r}, \omega) = -\frac{2VM}{\pi} \lim_{\delta \to 0_+} (\delta n_1(\mathbf{r}, \omega + i\delta) + \delta n_2(\mathbf{r}, \omega + i\delta) - \delta \leftrightarrow -\delta)$$

$$\delta n_1(\mathbf{r}, \omega) = \sum_{ij}\left(g_i(\mathbf{r}, \omega + i\delta) + \tilde{g}_i(\mathbf{r}, \omega + i\delta)e^{i\mathbf{Q}_x \cdot \mathbf{r}}\right)\left(g_j(\text{-}\mathbf{r}, \omega + i\delta) + \tilde{g}_j(\text{-}\mathbf{r}, \omega + i\delta)e^{-i\mathbf{Q}_x \cdot \mathbf{r}}\right)$$

$$\delta n_2(\mathbf{r}, \omega) = 2(1 + \cos(\mathbf{Q}_x \cdot \mathbf{r}))(f(\mathbf{r}, \omega + i\delta) + \tilde{f}(\mathbf{r}, \omega + i\delta))\left(f(\text{-}\mathbf{r}, \omega + i\delta) + \tilde{f}(\text{-}\mathbf{r}, \omega + i\delta)\right)$$

$\delta n(\mathbf{r}, \omega)$ defined in this way may be directly compared to the experimentally determined "cropped" QPI associated with a single impurity. In Fig. 2 we used $\xi^{-1} = 0, \Delta_{LRA} = 0.05$, (panel g) and $\Delta_{LRA} = 0.1$ (panel h).

The temperature dependence of anisotropy parameter results shown in Fig. 3g were obtained using a mean-field like ansatz for the magnetic correlation length and the mean-field order parameter:

$$\xi^{-1} = \xi_0^{-1}\sqrt{\frac{T-T_N}{T_S-T_N}}; \Delta_{SDW} = \Delta_0\sqrt{1-\frac{T}{T_N}}$$

with $T_N = 40K, T_S = 52K, \xi_0 = 20, \Delta_0 = 0.14$ and $\Delta_{LRA} = 0.052$.

To model the momentum space data we Fourier transform the real-space calculations and take the absolute value.

### V. Measuring experimental QPI and comparison with theory

The experimental QPI signal can be determined in one of two ways – it can either be determined from the FT of a single defect, or from the FT of a large area map that includes many defects. While the two procedures give similar results, they differ in some important respects. Figure S2 illustrates this difference. Shown in area map of NaFeAs at taken at 10meV conditions at 26K. This image is chosen since individual defects can be clearly distinguished from each other. Thus, it is relatively simple to crop around each defect and average together the QPI signal from all the cropped areas to generate the QPI signal associated with a single defect. The result of this procedure is shown in the inset of Fig. S2a. The corresponding Fourier transforms (FT) of Fig. S2a and inset S2a are shown in panels S2b and S2c respectively. The FTs have been cropped to half of the 1 Fe BZ of NaFeAs. As can be seen, the FTs have many similarities, but there are also several differences in the two FTs. In particular, the central "stripe" seen in the FT of the full real space image looks different in the FT of the average defect, where it shows up as two separated regions of intensity. The difference in the two procedures is largely due to the fact that when the FT is taken of an entire image with several defects, one is in effect adding together signals from defects that are distributed in space, each of which gives rise to a phase factor from the location of the defect. The sum of these phases is in general a strong k-dependent function that depends on the distribution of defects in space. In practice, this factor is mostly (but not fully) spherically symmetric. Thus, Fourier space difference plots of the QPI signal extracted in these two different ways are quite similar to each other (but not identical) as shown in inset S2b and inset S2c.

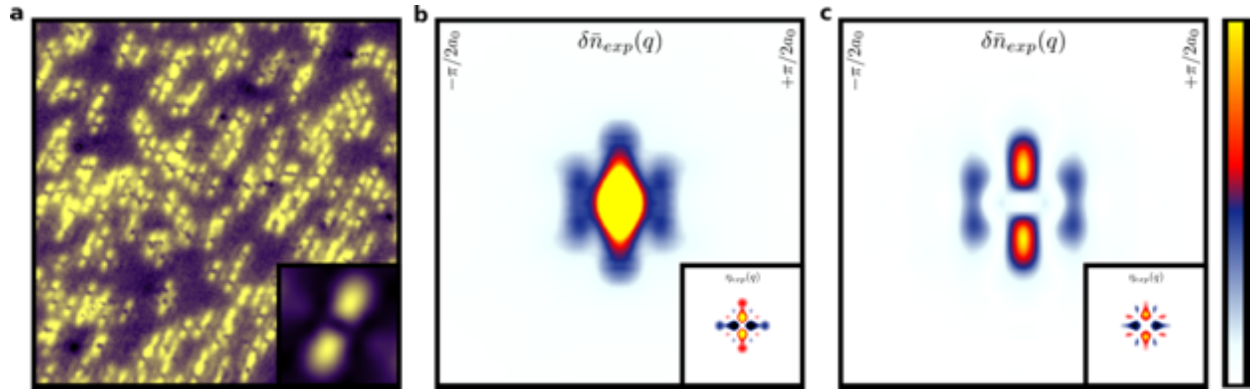

**Figure S2:** a, ~100nm real space spectroscopic image take at 10meV and at 26K. Inset 8.3nm image showing the average of all the defects in image a. b, Fourier transform of real space image a from $-\pi/2a_0$ to $\pi/2a_0$ with inset showing the resulting difference plot. c, Fourier transform of average defect image in inset a from $-\pi/2a_0$ to $\pi/2a_0$ with inset showing the resulting difference plot.

The theory for QPI that we (and others) use refers to the scattering pattern in k-space (or real space) generated by a single impurity, and thus the true comparison should be made to the average defect FT. Indeed, the low temperature QPI from theory (Fig. S3a) matches quite well with the single-defect experimental QPI pattern (Fig. S3b). The theory calculation is performed for a model as described above with parameters $\xi^{-1} = 0$ and $\Delta_{LRA} = 0.1$. The bright points along the $q_y$ directions (green arrows in Fig. S3a-b) as well as the outer features that run parallel to the center bright points (purple arrow in Fig. S3a-b) are both reproduced in theory. The center bright points along the $q_y$ direction (green arrow Fig. S3a-b) have a slightly different scattering vectors lengths with the experiment having a smaller vector length when compared to theory, falling within a $.03\pi/a_0$ range of each other. The outer features (purple arrow Fig. S3a-b) are father from the center in $q_x$ then what is seen in experiment, falling within a $.1\pi/a0$ range. While theory doesn't capture all the details seen in experimental QPI, it indeed captures the import scattering vectors from the band structure near the Fermi level as discussed in greater detail in ref. [1].

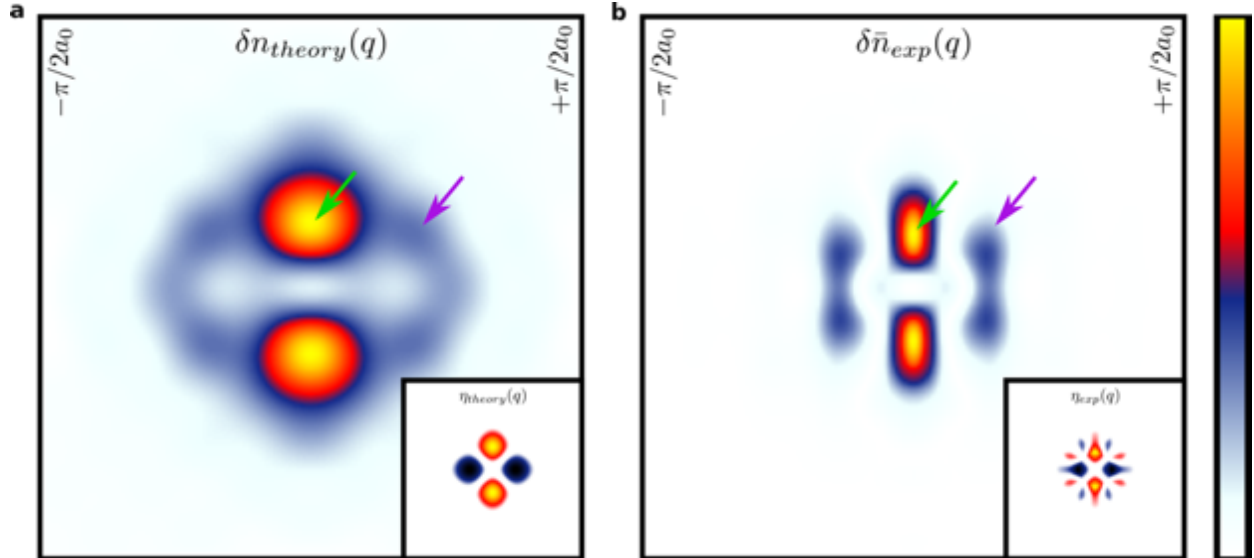

**Figure S3:** a, Theoretical QPI calculation from $-\pi/2a_0$ to $\pi/2a_0$ with inset showing the resulting difference plot. b, Fourier transform of average defect image in inset a from $-\pi/2a_0$ to $\pi/2a_0$ with inset showing the resulting difference plot.

### VI. Nematic Anisotropy Coupled to Induced Strain

Shown in Fig. S4 is a sequence of STS images taken over the same region of the parent NaFeAs sample at different values of the applied strain at 54K. The strain is varied starting from -200V going up to +200V and then reversing back to -200V. STS images are shown at three biases:

+10mV, +20mV, and +30mV. All images are obtained under the same tunneling conditions ($V_{set}$=-50mV, I=-100pA). Anisotropy in the images shows up as white streaks in the images that are oriented nearly vertically. The overall magnitude of the anisotropy is strongly reduced from its low temperature value in all the images. Considering figures S4a-e taken at a bias voltage of +10 mV, it is seen that the anisotropy is maximal at a strain voltage of +200V (Fig. S4c) while it is nearly absent at -200V (Fig S4a and e). The anisotropy is seen to be a continuous function of strain with no domains appearing at any strain value. The images also show no evidence for hysteresis. Similar behavior is seen in the +20mV STS scans (Fig. S4f-j) and +30mV STS scans (Fig. S4k-o). The overall magnitude of the anisotropy is small below -10mV and above +30 mV.

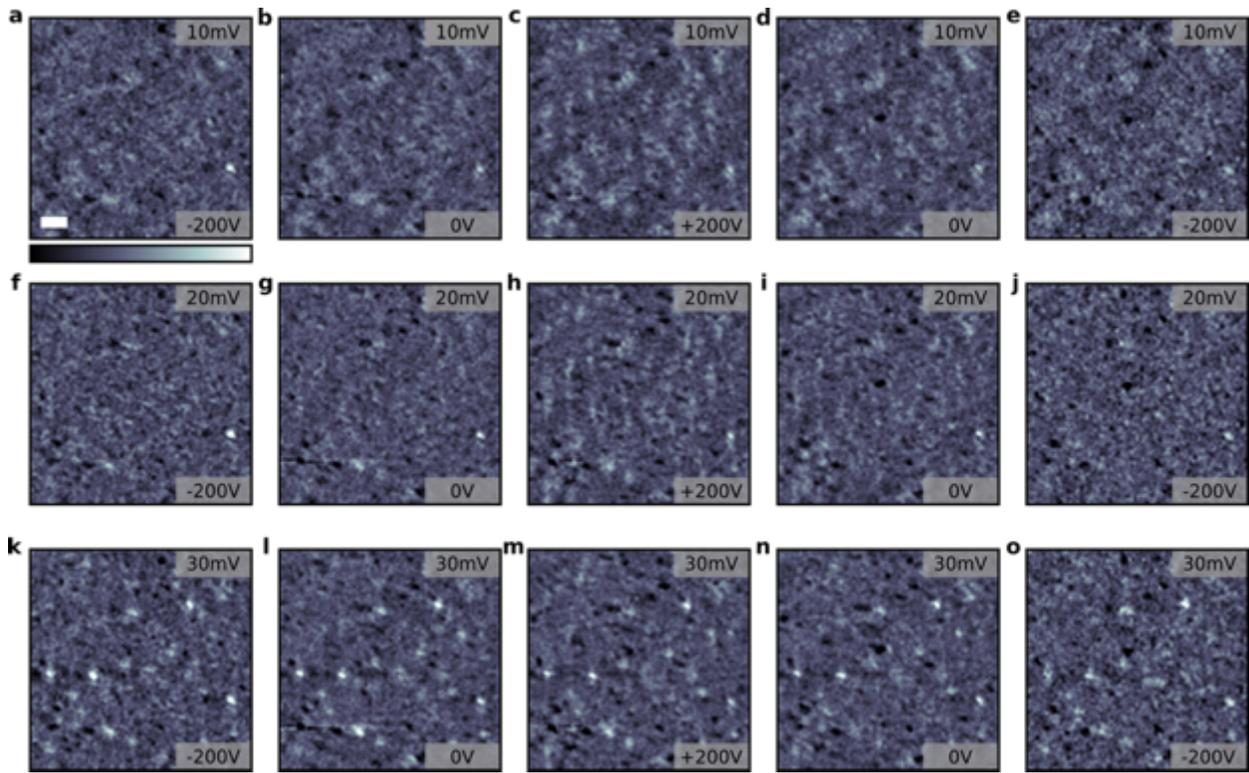

**Figure S4:** a-e, Real space, 10mV E-STS scans at 54K progressing from maximal applied compressive strain at -200V to maximal applied tensile strain at +200V and back to maximum applied compressive strain at -200V. f-j, 20 mV E-STS scans for same conditions as a-e. k-o, 30mV E-STS scans for same conditions as a-e.

### VII. Domain Wall Disappearance at $T_S$

In our temperature-dependent measurements shown in figure 3, we have displayed areas that are about 100 nm x 100 nm, and have observed that domain walls disappear precisely at $T_s$ in this region. To address whether this disappearance of the domain walls is true across the entire sample, we have performed the following additional measurements:

(a) To look at larger areas, we have scanned areas > 500 nm x 500 nm across the structural transition, which is at the limit of our STM scan range while keeping atomic registry with temperature. A subset of these images is shown in figure S5. The images in Fig. S5a-b show effectively the same area of the sample at 45K and 49K (below $T_S$). We can clearly see domain boundaries (>20) appear as lines on these images, and we also see interesting domain wall motion as a function of temperature just below $T_S$ (which is not seen at very low temperature). The image shown in figure S5c is taken above $T_S$ and it is clear that there are no domain walls in the figure. This extends the statistics of figure 3 to a much larger number of domain walls, and we indeed see that all the domain walls are absent above $T_S$.

(b) To get even better statistics on the presence of domain walls, we scan multiple areas of the sample at each temperature. The fine scan limit of our STM is between 1-1.5 μm (depending on temperature) and we can futher move around macroscopically on the sample to different locations with coarse motors. We have scanned (conservative estimate) about 80 μm² on the parent compound of NaFeAs across tens of samples at temperatures just below (temperatures that range from 3-7 K below $T_S$) and above (temperatures that range from 1-7 K above $T_S$) $T_S$. From our measurements just below $T_S$ we find that the average size of the domain is about 0.05 square microns. On the other hand, we have never seen any domains above $T_S$ in all of our measurements. If we assume a Poisson distribution of the density of domain walls, our observation of no domain walls implies that the probability that domain walls exist above $T_S$ (but we have missed them in all our measurements so far) is $< 10^{-6}$.

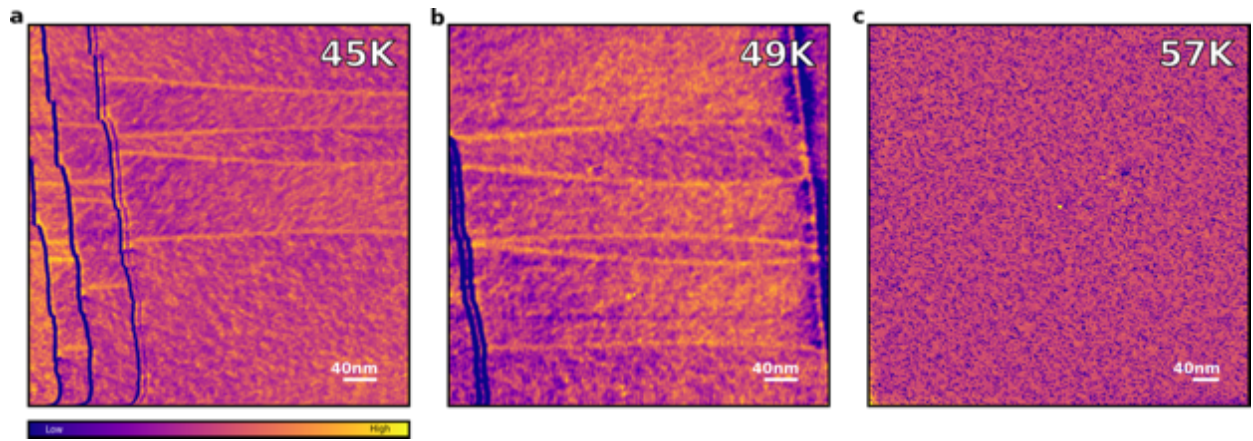

Figure S5: a-c, STM images at below (a,b) and above (c) $T_S$ showing the evolution of domain walls with temperature. Below $T_S$, numerous (>20) domain walls are observed, which are visualized as curved lines on these images. Domains are observed to shrink or grow depending on the orientation of their anisotropy. No domain walls are observed above $T_S$.

## VIII. Temperature dependence of anisotropy parameter

The temperature dependence of the anisotropy parameter in figure 3g is based on the STS maps in figures 3a-f. For temperatures below $T_S$, two domains are seen in the STS maps, and the data points shown in figure 3g are based on the domain that survives above $T_S$ as shown in figure S6a.

For the domain that disappears at $T_S$, we can analyze the anisotropy parameter as a function of temperature up to the highest temperature at which the domain is observed. The results of this analysis are shown in figure S6b as red crosses, together with the existing data points from figure 3g of the main text (open circles). We can see that the data for both domains lie almost identically on each other, indicating that the sharp drop in intensity is seen for both domain orientations near the magnetic transition. Additionally, the analysis of domains in both directions also removes any uncertainty due to anisotropic tip shape.

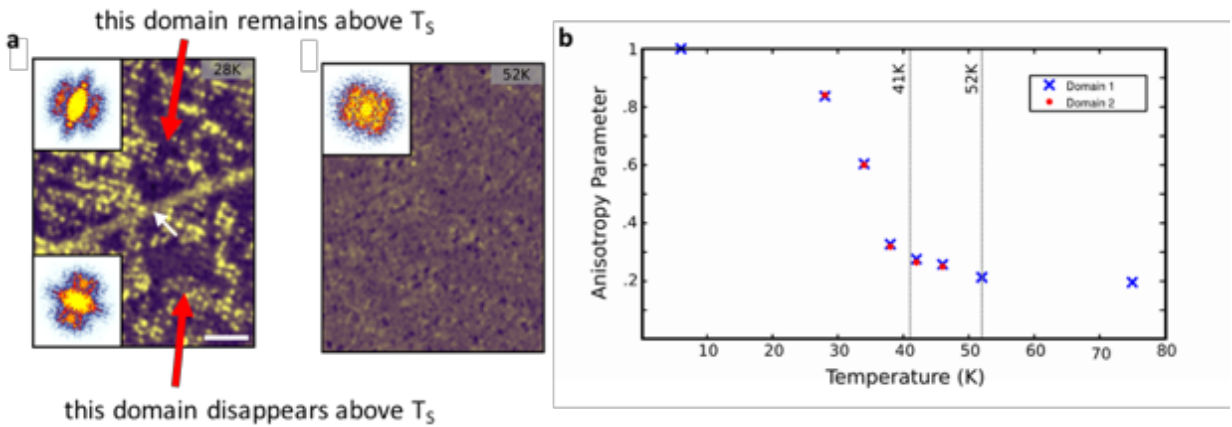

Figure S6: a STM images below (left) and above (right) T$_S$ (identical to figure 3a and 3f of the main text respectively) showing that one of the domains survives above T$_S$. b Plot of the anisotropy parameter extracted from each of the domains. The blue crosses correspond to the domain that survives above T$_S$ while the red dots correspond to the domain that disappears at T$_S$. Both data sets show a sharp drop in intensity near the magnetic transition temperature (~40 K) rather than the structural transition temperature (~ 51 K).